# Cooldown Measurements in a Standing Wave Thermoacoustic Refrigerator

R. C. Dhuley, M.D. Atrey

*Mechanical Engineering Department,
Indian Institute of Technology Bombay, Powai
Mumbai-400076*

*Thermoacoustic Refrigerators (TARs) use acoustic power to generate cold temperatures. Apart from the operating frequency and the mean temperature of the working medium, the charging pressure and the dynamic pressure in the TAR govern its attainable cold temperature. The effect of charging pressure on the dynamic pressure in a loudspeaker driven gas filled standing wave column has been well understood. The present work aims to investigate the effect of charging pressure on the cold end temperature of a standing wave TAR. The cold end temperature lift and the cooldown for several changing pressures are reported. The effect of vacuum around the cold end on the TAR performance is also presented.*

Key words: Thermoacoustic, standing, acoustic, refrigerator

## INTRODUCTION

Thermoacoustic Refrigerators (TARs) use the energy of sound waves to produce cold temperatures. The thermal interaction of the compressions and rarefactions present in a sound wave with a porous medium, sets up a heat flow across the porous medium. This flow can be employed to pump heat from one location of the porous medium to the other.

The operating frequency, the mean temperature and the charging pressure of working medium, and the dynamic pressure inside the TAR, are the main parameters on which the performance of a TAR depends. The dependency of the dynamic pressure on the charging pressure and the operating frequency, in loudspeaker driven standing wave columns has been described elsewhere [1]. The present work aims to understand the effect of charging pressure on the cold end temperature of a quarter-standing wave TAR. The effect of vacuum insulation around the cold end of the refrigerator is also investigated.

## QUARTER WAVE RESONATOR

The half wave (straight) resonators are quite easy to design for a given operating frequency, and to manufacture. However, they exhibit non-linear effects when they are driven at large amplitudes at their resonance frequency. The occurrence of such a non-linear effect - periodic shock waves, has been encountered in our previous investigations [1]. Such periodic shock waves deteriorate the acoustic amplitude inside the resonator and disturb the standing wave phasing. They are also detrimental to the acoustic driver. These non-linear effects can be minimized by means of a resonator that has non-uniform cross section over its length [2]. Such a resonator can be designed to resonate at the operating frequency in its quarter wave mode.

In the present work, commercial software DeltaEC [3] is used to design the quarter wave resonator. DeltaEC integrates one-dimensional thermoacoustic equations in a user-defined acoustic geometry, satisfying the given boundary conditions. Figure 1 shows the schematic of the quarter wave resonator.

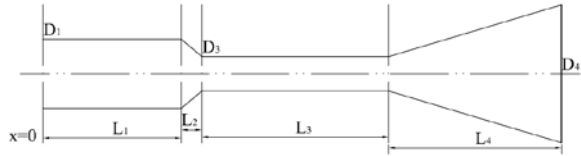

Figure 1. Schematic of a quarter wave resonator

The resonator consists of a larger diameter tube of diameter $D_1$ and length $L_1$ (the stack holder), a conical reducer of length $L_2$, a smaller diameter tube of diameter $D_3$ and length $L_3$, and a conical buffer volume of length $L_4$ in specified order. The larger diameter of the conical buffer is $D_4$. The acoustic driver is located at the beginning of the larger diameter tube at x=0. All the geometric parameters shown in the Figure 1, except $L_3$, are also assumed. The rigid termination at the extreme end of the buffer volume is a region of zero oscillatory velocity, hence a location of infinite acoustic impedance. In order to enforce resonance at the chosen operating frequency, the oscillatory pressure and velocity at x=0 are kept in phase. Table 1 shows the values of all the geometric dimensions and the resultant value of $L_3$ for resonance at 400 Hz in Helium gas.

Table 1. Resonator dimensions

| Length | Value | Diameter | Value |
|---|---|---|---|
| $L_1$ | 0.1 m | $D_1$ | 0.032 m |
| $L_2$ | 0.02 m | $D_2$ | 0.0126 m |
| $L_3$ | 0.1846 m | $D_3$ | 0.12 m |
| $L_4$ | 0.1 m | | |

**EXPERIMENTAL SETUP**

To do the proposed study, a standing wave TAR driven by an acoustic driver is developed. The acoustic driver is constructed from an electro-dynamic motor of a commercially available loudspeaker. The driver is constructed suitably in order to withstand the high charging pressure of the working gas. To satisfy the requirement of high strength to sustain the large dynamic pressure that is generated during operation, a new voice coil is fabricated. The former of the new voice coil is made from Delrin (as against the paper voice coil in the commercial package), which ensures light weight and high strength. Super-insulated copper wire is used for winding. A thin rubber sheet is used to suspend the voice coil in the magnetic flux gap. This acoustic driver assembly is contained in a stainless steel enclosure which has a provision for attaching the resonator and the vacuum jacket.

For the cooldown measurements, a spiral stack is manufactured. This geometry is chosen owing to the simplicity in its construction procedure. The stack is made by gluing 0.3 mm Nylon fishing lines onto a 0.18 mm thick Mylar film, and then slowly winding it into a roll. The spacing between layers of Mylar film corresponds to $2.5\delta_k$ which is close to its optimal value of $2\delta_k$ [2] ($\delta_k$ is the thermal penetration depth). The porosity of the stack is 0.53, the length is 100 mm ($L_1$) and its diameter is 32 mm ($D_1$).

The stack holder and the smaller diameter tube of resonator are manufactured from Delrin, principally to ensure low thermal conductivity, high strength and light weight. The conical part of the resonator is incorporated into the smaller diameter tube. A provision to attach a dynamic pressure sensor is also made in the conical part of the resonator. All dimensions of both the components are already given in Table 1.

A thin copper ring is sandwiched between the two resonator components (near the cold end of the stack) to provide a surface to mount the temperature sensor. The inner surface of the ring is in contact with the working gas inside while the temperature sensor can be mounted on its outer surface. Due to large cooldown time of this copper ring (because of its large thermal mass), it was decided to measure the temperature of the gas directly. For this purpose, a small hole is drilled in the copper ring through which a thermocouple can be inserted into the resonator.

The buffer volume at the end of the resonator is made by rolling a 1 mm thick stainless steel sheet in the shape of a cone and welding it along its slant length. A circular sheet of stainless sheet is welded to the larger diameter end of the cone, which terminates the resonator. All the components of the resonator are provided with connecting flanges so that they can be fastened by nut-bolts. This facilitates easy assembly and dismantling of the resonator components.

An aluminium vacuum jacket is also manufactured to study the effect of vacuum insulation on the cooldown rate and the cold end temperature of the stack. This jacket encloses the entire resonator and has a provision for electrical feed-throughs for temperature and dynamic pressure measurements.

A schematic of the TAR assembly is shown in Figure 2.

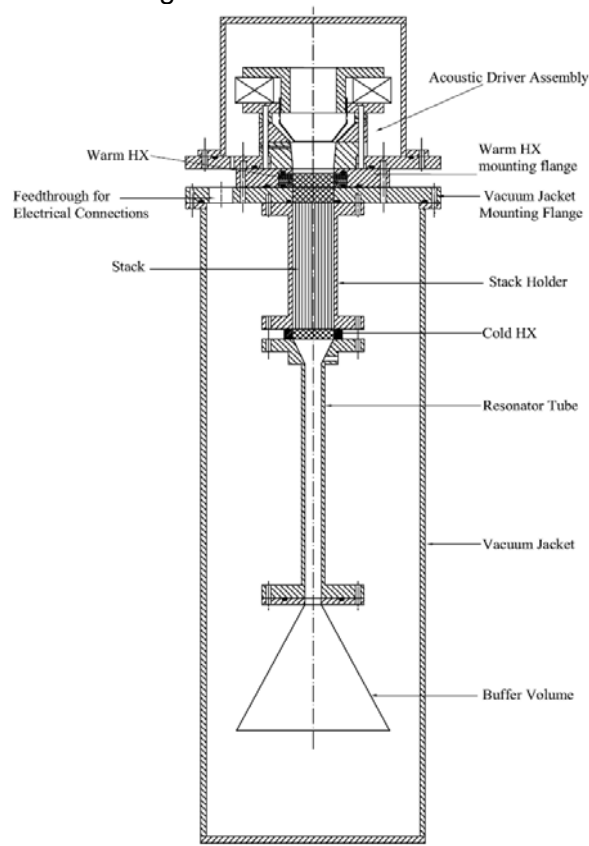

Figure 2. Schematic of the TAR setup

An AC power supply is used to drive the TAR. This power supply has provisions to vary the supply voltage and the frequency independently. The input power is monitored by means of a digital power meter. A piezoresistive transducer is used to measure the dynamic pressure generated by the acoustic driver. The output of the pressure sensor appears on an oscilloscope in the form of a sinusoidal waveform. The magnitude of the dynamic pressure as well as resonance can be determined from this waveform. The cold end temperature of the TAR is measured by means of a T-type (copper-constantine) thermocouple. During operation, the thermocouple readings are logged into a DataTaker (a data acquision and logging system) which can also be viewed on a PC in realtime. The TAR setup (w/o the vacuum jacket) under operation along with the allied instrumentation is shown in Figure 3.

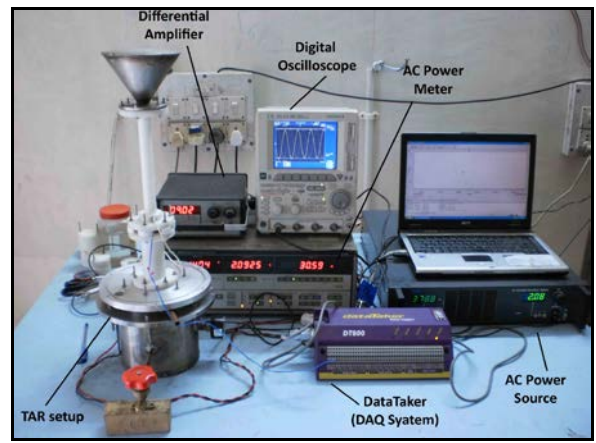

Figure 3. The TAR under operation

## RESULTS AND DISCUSSION

### a) Resonance frequency measurement

Prior to the cooldown measurements, the resonance frequency of the new resonator is determined experimentally. After the assembly, the setup is evacuated to remove the air present inside and then charged with 10 bar Helium gas. The acoustic driver is excited with a small voltage (~ 2-3 V) at 400 Hz frequency. A pressure wave is observed on the oscilloscope. Keeping the voltage constant, the frequency is slowly varied and the change in the amplitude of the pressure wave is studied. At the acoustic resonance, the magnitude of the standing pressure wave attains a maximum value. The magnitude starts to fall down as soon as there is a resonance cross-over in any direction. The resonance frequency can also be determined from the current reading of the power-meter. At resonance, the acoustic impedance of the resonator becomes maximum (and real) and hence, the TAR draws a minimum current for a given value of input voltage. Thus, the resonance is characterized by a trough in a current vs. frequency curve of the refrigerator. (However, it is to be noted that the current also attains a minimum value at the mechanical resonance of the acoustic driver. But at this resonance, the driver generates negligible dynamic pressure).

The resonance frequency is determined to be 384 Hz. As compared to the theoretical design of 400 Hz, the match is fairly good. The difference can be accounted for by the increase in overall length of the resonator due to the copper ring that is sandwiched between the stack holder and the smaller diameter tube. Similar measurements are done for different values of charging pressures. It is observed that the resonance frequency is almost independent of the changing pressure.

**b) Cooldown measurement**

After determining the resonance frequency, the input frequency is kept constant at 384 Hz. Thereafter, the voltage input is slowly increased till the input power reaches the desired value. It is observed that the temperature of one end of the stack begins to fall while that of the other begins to rise. In the present work, the rise in temperature of the heating end of the stack could not be measured due to an intrinsic difficulty in mounting a temperature sensor at that location. Due to absence of the ambient heat exchanger and the mechanism to remove the heat generated in acoustic driver, the duration of each trial is deliberately kept to 20 minutes. This is done to prevent overheating and eventual burnout of the acoustic driver.

Experiments are carried out at various charging pressure *viz.* 10 bar, 8 bar, 6 bar and 4 bar with same 20 W input power. Figure 4 shows the fall in temperature of the cold end of the stack with time for different charging pressures. The vacuum jacket was not used for these measurements.

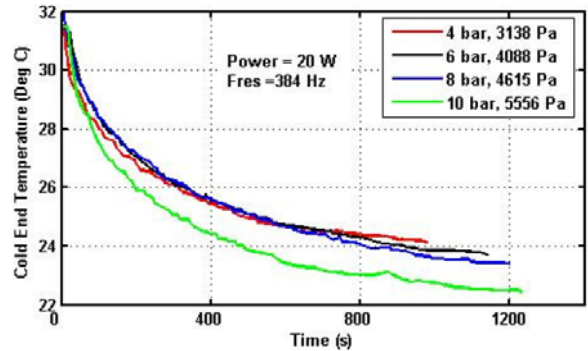

Figure 4. Cooldown curves for different charging pressure

As concluded in [1], for the same input power, higher dynamic pressure is generated in the resonator when the charging pressure is more. This is also clear from the legend in Figure 4, where the former quantity is the charging pressure and the later represents the corresponding dynamic pressure that is generated.

Referring to Figure 4, the cooldown is rapid in the initial stage of operation and as time passes, the cooldown rate becomes slow. The initial rate of temperature fall is almost same for various charging pressures, but the final steady state temperature is lower when the charging pressure is more. After about 20 minutes, the cold end temperature becomes almost stable for all the values of charging pressures. The experiment at 4 bar could not be run till 20 minutes due to a loose connection of a wire that powered the acoustic driver.

The steady state cold end temperatures for different charging pressures are shown in Figure 5.

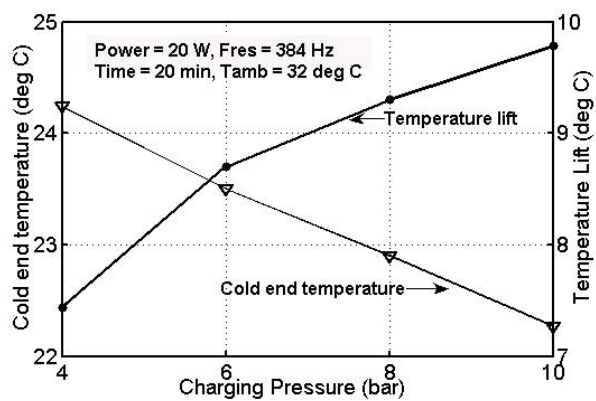

Figure 5. Steady state cold temperature

The corresponding temperature lift from the ambient is also shown in Figure 5. As mentioned above, the cold end temperature is less when the charging pressure is more. This is so because higher dynamic pressure is generated when the charging pressure is more. In course of these experiments, the minimum cold end temperature of 22 $^0$C is reached at 10 bar charging pressure. The corresponding lift from ambient is 10 $^0$C.

### c) Cooldown measurement in vacuum

In the experiments described above, no insulation around the cold end of the stack (and the copper ring) is used. As a result, there are external heat leaks into the refrigerator. To minimise the heat leaks due to conduction and convection, vacuum insulation is used around the entire resonator. An aluminium vacuum jacket is used to enclose the entire resonator along with the stack and the copper ring.

The cooldown measurement with the vacuum jacket is carried out at 10 bar Helium and the result is compared with that obtained in absence of vacuum. The test is carried out for 20 minutes duration with the same power input of 20 W. Figure 6 shows the cooldown curves when the TAR is run in absence and in presence of vacuum ($10^{-3}$ mbar) around the cold end.

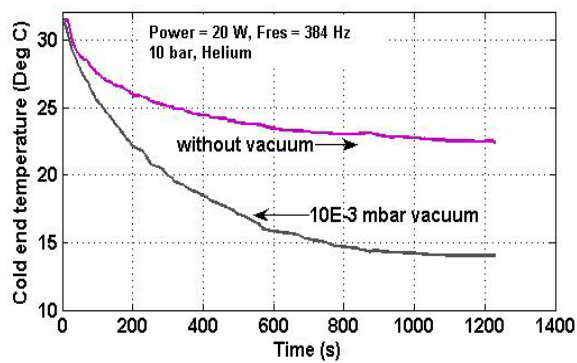

Figure 6. Cooldown curves in absence and in presence of vacuum

As seen in Figure 6, there is a significant improvement in the results. The initial rate of cooldown in both the cases is almost same. As temperature starts to fall, after about 60 sec, the cooldown rate in absence of vacuum starts to diminish due to external heat leak. After about 500 sec, the cold end temperature is close to 23 $^0$C. The rate of temperature fall almost becomes steady. After 20 minutes of operation, the cold end temperature stabilises at 22 $^0$C. However in vacuum, the cooldown rate is much faster. The temperature fall becomes slow after about 800 sec when the cold end temperature is close to 15 $^0$C. After 20 minutes, the cold end attains a temperature of 14 $^0$C, which is a lift of 18 $^0$C from the ambient temperature.

## CONCLUSIONS

The cooldown measurements on a standing wave TAR have been presented in this paper. Due to inadequate design of the acoustic driver, the experiments are carried out at low dynamic pressures. It is observed that higher charging pressure induces larger dynamic pressure at same input power and hence, results in lower no load temperature. The vacuum insulation around the cold end of the TAR minimizes the heat leak from the ambient to the cold end. As a result, there is a significant improvement in the cooldown rate and the steady state no load temperature of the TAR. For same operating conditions, the TAR attained 14 $^0$C in presence in vacuum as compared to 22 $^0$C in its absence.

Much higher dynamic pressures are needed for better performance of the TAR and to attain lower no load temperatures. To do this, an efficiently designed driver with a powerful motor is required. Further, to prevent driver burnout due to resistive heating during experiments, a mechanism to remove heat from the driver is essential. During present investigations, the temperature of the hot end of the stack rose well above the ambient due to heat pumping across the stack. To ensure proper cooldown of the stack, this heats needs to be removed from the TAR. A well designed heat exchanger at the hot end of the stack can serve both the functions.